\documentclass[prl,aps,twocolumn,superscriptaddress]{revtex4-1}

\usepackage{amsmath}
\usepackage{amssymb}
\usepackage{graphicx}
\usepackage{threeparttable,booktabs}
\usepackage{yhmath}
\usepackage{enumerate}

\usepackage[utf8]{inputenc}
\usepackage{bm}% bold math
\usepackage{appendix}

\usepackage{bbm}

\graphicspath{{plots/}}

\newcommand{\F}{\mathcal{F}}

\newcommand{\Sc}{\mathcal{S}}
\newcommand{\R}{{\mathbb R}}
\newcommand{\Vr}{\mathcal{R}}
\newcommand{\Hc}{{\mathcal H}}

\renewcommand{\L}{\mathcal{L}}
\newcommand{\N}{\mathcal{N}}

\newcommand{\Cov}{\text{Cov}}
\newcommand{\Var}{\text{Var}}

\newcommand{\bu} {{\bf u}}

\newcommand{\bs} {{\bf s}}

\newcommand{\bsigma} {\bm{\sigma}}
\newcommand{\m}{{\bf m}}
\newcommand{\n}{{\bf n}}
\newcommand{\q}{q}
\newcommand{\bq}{\bm{q}}

\newcommand\egaldef{\stackrel{\mbox{\upshape\tiny def}}{=}}

\newcommand\1{\leavevmode\hbox{\rm \small1\kern-0.35em\normalsize1}}

\newcommand\EE{\mathbb{E}}
\def\DD{\displaystyle}
\DeclareMathOperator*{\lra}{\longrightarrow}
\DeclareMathOperator*{\msim}{\sim}

\DeclareMathOperator*{\argmin}{argmin}

\begin{document}

\title{Exact Training of Restricted Boltzmann Machines on Intrinsically Low Dimensional Data}
\author{A. Decelle}
\affiliation{LISN, AO team, B\^at 660 Universit\'e Paris-Saclay, Orsay Cedex 91405}
\affiliation{Departamento de Física Téorica I, Universidad Complutense, 28040 Madrid, Spain}
\author{C. Furtlehner}
\affiliation{Inria Saclay - Tau team, B\^at 660 Universit\'e Paris-Saclay, Orsay Cedex 91405}
\affiliation{LISN, AO team, B\^at 660 Universit\'e Paris-Saclay, Orsay Cedex 91405}

\begin{abstract}
  The restricted Boltzmann machine is a basic machine learning tool able, in principle, to model the distribution of some arbitrary dataset.
  Its standard training procedure appears however delicate and obscure in many respects.     
  We bring some new insights to it by considering the situation where the data have low intrinsic dimension,
  offering the possibility of an exact treatment and revealing a fundamental failure of the standard training procedure.
  The reasons for this failure \textemdash~like the occurrence of first-order phase transitions during training~\textemdash \ are clarified thanks to a Coulomb interactions
  reformulation of the model. In addition a convex relaxation
  of the original optimization problem is formulated thereby resulting in a unique solution,
  obtained in precise numerical form on $d=1,2$ study cases, while a constrained linear regression solution can be conjectured on the basis of an information theory argument.
\end{abstract}

\maketitle

Recent advances in machine learning (ML) pervade now many other scientific domains including physics by providing
new powerful data analysis tools in addition to traditional statistical ones. The restricted Boltzmann machine (RBM) could be considered as one of these when already
a large spectrum of possible uses has been proposed in physics~\cite{ToMe,CaTr,NoDaYaIm,MeCaCa,TuCoMo}. 
Introduced more than three decades ago~\cite{Smolensky}, the RBM played an important role in early developments of deep learning~\cite{HiSa}.
It is a special case of  generative models~\cite{GoPoMiXuWaOzCoBe,KiWe,SaHi}
\begin{figure}[ht]
\centerline{\resizebox*{0.65\columnwidth}{!}{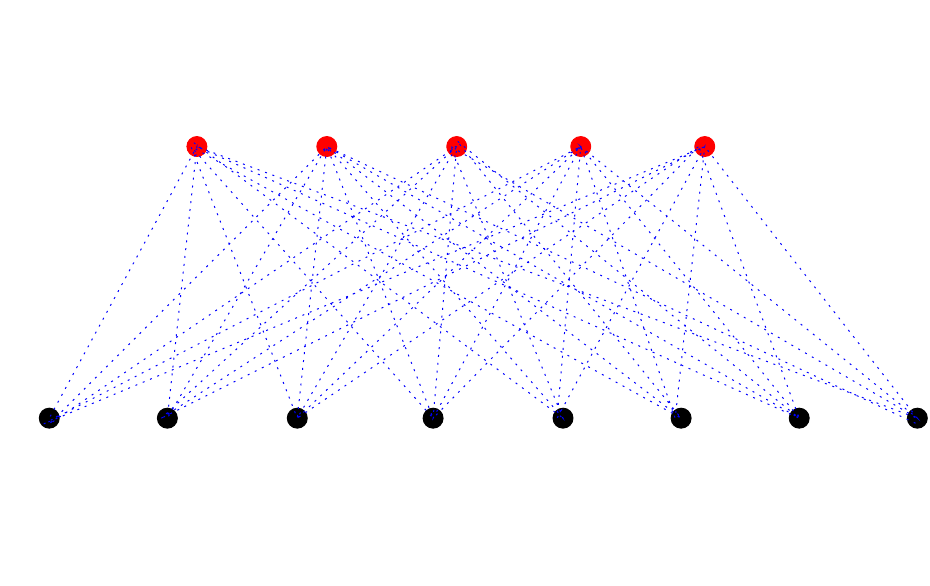}
\resizebox*{0.34\columnwidth}{!}{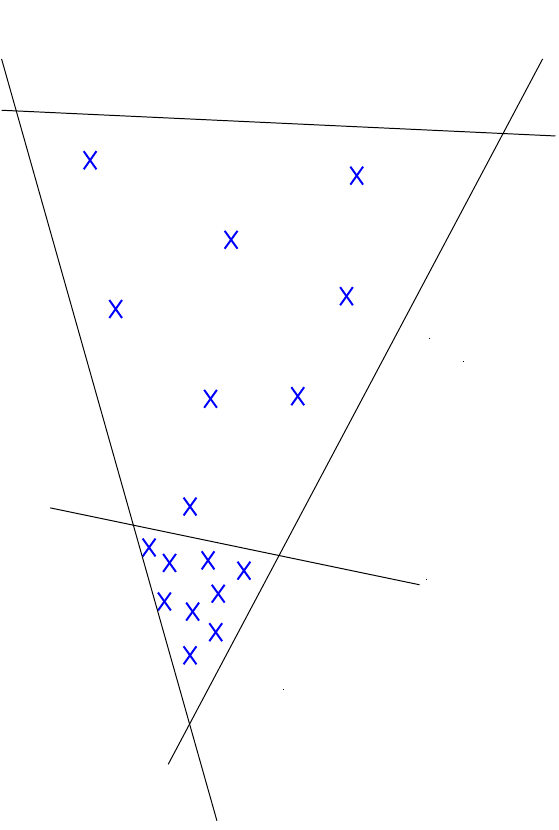}}
\caption{\label{fig:rbm} Bipartite structure of the RBM (left).
  Hyperplanes defined by the weight vectors and bias associated with each hidden variable can delimit fixed density regions in input space (right).}
\end{figure}
that remains very popular thanks to its simplicity and effectiveness when
applied to moderately high dimensional data~\cite{HjCaSaAlAdPl,HuHuPeHaLiLvGuGuLi,YeDeOnMaTaMoFuPaJa}. 
It is a two-layers undirected neural network that represents the data in the form of a Gibbs distribution of visible and latent variables (see Fig.~\ref{fig:rbm}): 
\begin{equation}
    p(\bs,\bsigma) = \frac{1}{Z[\Theta]}\exp\Bigl(\sum_{i,j} s_i W_{ij} \sigma_j - \sum_{i=1}^{N_v} \eta_i s_i - \sum_{j=1}^{N_h} \theta_j \sigma_j\Bigr).
    \label{eq:rbm}
\end{equation}
The former noted $\bs = \{s_i,i=1\ldots N_v\}$ 
correspond to explicit representations of the data while the latter noted ${\bm \sigma} = \{\sigma_j,j=1\ldots N_h\}$ 
are there to build arbitrary dependencies among the visible units. They play the role of an interacting field among visible nodes. 
While many different types of variables can be considered, we take here spin variables $s_i,\sigma_j \in \{-1,1\}$ for definiteness. $\Theta = (W,\bm{\eta},\bm{\theta})$ are the parameters,
$W$ being the weight matrix, $\bm{\eta}$ and $\bm{\theta}$  are local field vectors
called respectively visible and hidden biases. Each weight vector associated with a given hidden unit and its corresponding bias
defines an hyperplane partitioning the visible space into two regions corresponding to the hidden unit being activated or not (see Fig.~\ref{fig:rbm}).  
$Z[\Theta]$ is the partition function of the system. The joint distribution between visible
variables is then obtained by summing over hidden ones.
Learning the RBM amounts to find $\Theta$ such that generated data obtained by sampling this distribution 
should be statistically similar to the training data. The standard method to infer the parameters is to maximize the log-likelihood (LL) of the model
\begin{align}\label{eq:LL}
  \L[\Theta] = \sum_j \langle \log\cosh&\bigl(\sum_i W_{ij} s_i - \theta_j\bigr)\rangle_{\rm Data} \nonumber\\[0.2cm]
  &-\sum_i \eta_i \langle s_i \rangle_{\rm Data}- \log\bigl(Z[\Theta]\bigr),
\end{align}
with $\langle\rangle_{\rm Data}$ denoting the average over training data.
This is a nontrivial optimization problem in two respects: it is nonconvex and the loss function $-\L[\Theta]$ is difficult to estimate because $\log\bigl(Z[\Theta]\bigr)$ 
is not tractable. Nevertheless, the gradient $\nabla_\Theta\L[\Theta]$  can  be written in terms of simple response functions of the RBM.
These can be estimated approximately via Monte Carlo methods, leading to various algorithms called contrastive divergence~\cite{Hinton_CD} with possible
refinements~\cite{Tieleman,FiIg}. 
 
The similarity of the RBM with disordered spin systems has raised a lot of interest in statistical physics. Mean-field-based training algorithms  
and analyses have been proposed~\cite{GaTrKr,HuTo,TaYa,Mezard}, a mapping with the Hopfield model has been found in~\cite{BaBeSaCo},
retrieval capacity has been characterized in~\cite{BaGaSoTa,BaGaSoTa2} and compositional mechanisms analyzed in~\cite{Agliari,MoTu} 
(see more recent references, e.g. in~\cite{DeFu2020}).   

In previous works~\cite{DeFiFu2017,DeFiFu2018} we studied to what extent the learning process of the RBM is reflected in the spectral dynamics of the weight matrix: 
a certain number of modes, corresponding to principal modes of the data, emerge from a Marchenko-Pastur bulk at initialization and condense to build up a structured ferromagnetic phase.
Here we focus on the latter and most difficult stage and show that the two main difficulties (nontractability and nonconvexity) of the training can be
addressed in the special case, where a flat intrinsic space of low dimension has been identified in the first stage.
\paragraph*{Effective theory in the ferromagnetic phase.}%\label{sec:effective_th}
\textemdash~Let us first disentangle the contribution of the collective modes corresponding to the information stored from the data (the ferromagnetic and difficult part) 
from the other degrees of freedom corresponding to the noise (the paramagnetic and easy part).
After summing over the hidden variables in~(\ref{eq:rbm}) the visible distribution reads 
\begin{equation}\label{eq:Prbm}
P[\bs\vert\Theta] = \frac{1}{Z[\Theta]}\exp\Bigl[\sum_{j=1}^{N_h}\log\cosh\Bigl(\sum_{i=1}^{N_v} W_{ij}s_i-\theta_j\Bigr)-\sum_i\eta_i s_i\Bigr]
\end{equation}
As in~\cite{DeFiFu2018} the weight matrix is expressed via its singular value decomposition (SVD)
\[
W_{ij} = \sum_{\alpha=1}^{\min(N_v,N_h)} w_\alpha u_i^\alpha v_j^\alpha,
\]
with $w_\alpha$, ${\bm u}^\alpha$ and ${\bm v}^\alpha$ representing, respectively, the singular values and left and right singular vectors. 
Assume that some modes $\alpha\in\{1,\ldots d\}$ have condensed along a magnetization vector denoted $\m=(m_1,\ldots,m_d)$, i.e. that
$s_\alpha = m_\alpha = {\cal O}(1)$, with by definition
\[
s_\alpha \egaldef  \frac{1}{\sqrt{N_v}}\sum_{i=1}^{N_v} s_i u_i^\alpha.
\]
For a RBM trained on some data, $d$ would represent their intrinsic dimension at least locally. The corresponding modes $u_i^\alpha$ can, in principle, be
obtained directly from the SVD of the data or emerge naturally from the linear regime of the learning process described in~\cite{DeFiFu2018}.
These magnetization constraints define a canonical statistical ensemble. We look for a change of variables 
$\bs\longrightarrow (\m,\bs^\perp)$, where the original spin variables are replaced by a set of $d$ continuous variables
and ${\cal N}[\m]$ transverse weakly interacting spin variables. $\N[\m]$ is related to the configurational entropy  
per spin ${\Sc}[\m] = \frac{\N[\m]}{N_v}\log(2)$ under these constraints.
Thanks to a large deviation argument ${\Sc}[\m]$   is the Legendre transform of (see SM, Appendix~\ref{app:Fpara})
\[
\Phi[\bm{\mu}] = \frac{1}{N_v}\sum_i\log\cosh\Bigl(\sqrt{N_v}\sum_{\alpha=1}^d u_i^\alpha\mu_\alpha\Bigr),
\]
with $\bm{\mu}[\m]$ given implicitly  by the constraints~\footnote{Note that practically speaking  we use  finite $N_v$ estimates of $\Phi$ and $m_\alpha$
so that the preceding relation is in fact valid up to some ${\cal O}(1/\sqrt{N_v})$ corrections w.r.t. limit defined 
by some hypothetical $p_{\bm u}$ when $N_v\to\infty$.}
\begin{equation}\label{eq:malpha}
m_\alpha = \frac{1}{\sqrt{N_v}}\sum_{i=1}^{N_v} u_i^\alpha\tanh\Bigl(\sqrt{N_v}\sum_{\beta=1}^d u_i^\beta \mu_\beta\bigr),\ \alpha=1,\ldots d.
\end{equation}
%\subsection{Effective Hamiltonian}
Given a condensed magnetization vector $\m$, there remains ${\cal N}[\m]$
interacting degrees of freedom represented by spin variables denoted $\{s^\perp_1,\ldots,s^\perp_{{\cal N}[\m]}\}$.
With help of this new set of visible variables the partition function takes the form of a $d$-dimensional integral 
\begin{equation}\label{eq:ZTheta}
Z[\Theta]  = \int_{{\mathcal D}\subset [-1,1]^d} d^d\m\ e^{-N_v\F[\m\vert\Theta]},
\end{equation}
where the canonical free energy $\F[\m\vert\Theta] = \F^\parallel[\m\vert\Theta]+\F^\perp[\m\vert\Theta]$
is decomposed into two contributions coming respectively from the condensed modes and the transverse fluctuations (See SM, Appendix~\ref{app:FLT}):
\begin{align}
\F^\parallel[\m\vert\Theta] &= -\Sc[\m]+\sum_{\alpha=1}^d\eta_\alpha m_\alpha-V[\m\vert\Theta],\label{eq:Fpara}\\[0.2cm]  
\F^\perp[\m\vert\Theta] &= -\frac{1}{N_v}\log\Bigl(\frac{1}{2^{\N[\m]}}\sum_{\bs^\perp} e^{-\Hc_{\rm eff}[\bs^\perp\vert \m,\Theta]}\Bigr),\label{eq:Fperp}
\end{align}
($\eta_\alpha \egaldef \frac{1}{\sqrt{N_v}}\sum_i \eta_i u_i^\alpha$) which are respectively associated  with a potential function for the magnetizations 
\begin{equation}\label{eq:Vm}
V[\m\vert\Theta] = \frac{1}{N_v}\sum_{j=1}^{N_h}\log\cosh\bigl(\sqrt{N_v}\sum_{\alpha=1}^d w_\alpha m_\alpha v_j^\alpha -\theta_j\bigr),
\end{equation}
and an effective Hamiltonian $\Hc_{\rm eff}$ for the transverse degrees of freedom given in the form of a disordered Ising model
of $\N[\m]$ spins with paramagnetic-like state of order defined for each $\m$ (see SM, Appendix~\ref{app:EffH}).
The default entropy ($\N[\m]\log(2)$) of the transverse variables is assigned by convenience to $\F^\parallel$ so that $\F^\perp$ vanishes when $\Hc_{\rm eff}=0$.
In the following we focus on the dominant aspects of the training process resulting from the expression $\F^\parallel$.We leave aside
specific training problems associated with the transverse fluctuations, like e.g. the emergence of spurious modes,
which will be analyzed elsewhere in detail thanks to this effective Hamiltonian formalism.
\paragraph*{Coulomb formulation and linear regression.}%\label{sec:Coulomb}
\textemdash~The potential term in $\F^\parallel$,
which acts on the magnetization $\m$ representing here the position of a particle in a $d$-dimensional space, can be re-written as (See SM Appendix~\ref{app:Coulomb})
\begin{equation}\label{eq:VCoulomb}
V[\m\vert\Theta] = \int d\n dz\ \q(\n,z) \vert\n^T\m- z\vert,
\end{equation}
after introducing in the space $O(d)\times\R$, the density  
\begin{equation}\label{eq:qntheta}
\q(\n,z) = \frac{2}{N_v}\sum_{j=1}^{N_h} \nu_j\delta_{\nu_j}\Bigl(z-\frac{\theta_j}{\nu_j}\Bigr)\delta(\n-\n_j)\ge 0,
\end{equation}
of latent features, $\delta_\nu(x) = \frac{\nu}{2}\bigl[1-\tanh^2(\nu x)\bigr]$ being a ``smoothed" delta function of width $\nu^{-1}$, with
%\begin{minipage}{0.5\columnwidth}
\begin{align}
  \nu_j &= \sqrt{N_v\sum_{\alpha=1}^d w_\alpha^2{v_j^\alpha}^2}\label{eq:nuj} \\
%\end{equation}
%\end{minipage}
%\begin{minipage}{0.49\columnwidth}
%\begin{equation}
  n_j^\alpha &= \frac{\sqrt{N_v}}{\nu_j} w_\alpha v_j^\alpha\label{eq:nj}
\end{align}
%\end{minipage}
The kernel $\vert\n_j^T\m- z\vert$ represents the Coulomb potential exerted by a uniformly charged hyperplane, defined by its normal vector $\n$
and its distance $z$ to the origin, on a charge located at $\m$. As a result, each feature $j$ corresponds also to 
a charged hyperplane of normal vector $\n_j$,  offset $z_j=\theta_j/\nu_j$ and finite thickness $\nu_j^{-1}$.
At this point let us remark that the $w_\alpha$ control through~(\ref{eq:nuj}) both
the strength of the Coulomb interaction via~(\ref{eq:VCoulomb},\ref{eq:qntheta})
and the charged hyperplanes thickness; the right singular vectors projections
$v_j^\alpha$ control on their side the orientation of these hyperplanes in the intrinsic space through~(\ref{eq:nj}).
Note that the visible bias vector $\bm{\eta}$ is equivalent to some surface charge placed at the edge of the domain of $\m$
and can be incorporated into $q(\n,z)$. The log-likelihood of the RBM has then three terms 
\[
\L[\Theta] = -{\mathbb E}_{\hat p}\bigl[V[\m\vert\Theta]+\F^\perp[\m\vert\Theta]\bigr] - \log\bigl(Z[\Theta]\bigr),
\]
where $\log\bigl(Z[\Theta]\bigr)$ is a complex self-interaction of the charged hyperplanes among each other; ${\mathbb E}_{\hat p}\bigl[\F^\perp[\m\vert\Theta]\bigr]$ is in principle small, especially if there is
no transverse bias; finally,
\begin{equation}\label{eq:Coulomb}
{\mathbb E}_{\hat p}\bigl[V[\m\vert\Theta]\bigr] = \int d\m d\n dz\ \hat p(\m) \vert\n^T\m- z\vert q(\n,z),
\end{equation}
takes the form of a repulsive Coulomb interaction between training data points represented by the empirical distribution $\hat p(\m)$, and positively charged hyperplanes. 
It corresponds to a slight extension of the RBM model in terms of more general activation function (encompassing RELU~\cite{NaHi} for instance and similar to~\cite{PiLiIh}),
where each feature contribution in~(\ref{eq:Prbm}) comes with a non-negative weight $q_j$ to be optimized,
while the features themselves defined by the pairs $(\n_j,\theta_j)$ are predefined. 
This formulation introduced here at first  in a theoretical perspective to understand the RBM, can also be used in practice when the intrinsic space is identified in advance.
Then letting $w_\beta=0$ for $\beta>d$ results in  $\F^\perp$ independent of $\Theta$ and
the optimization of $\L[\Theta]$ (w.r.t. the features weights $\q(\n,z)$) becomes
convex, this ``Coulomb" formulation being in the exponential family.  
As a result the optimal solution can be obtained with good numerical precision thanks to a natural gradient ascent~\cite{Amari} 
following the geodesics of the Fisher metric (See SM, Appendix~\ref{app:exp}), the complexity being ${\cal O}(N_f^3+N_f^2\times N_p^d)$ in the number of predefined features $N_f$ and of points
$N_p^d$ needed to compute $Z[\Theta]$ (and its derivatives) through~(\ref{eq:ZTheta}). Typically this remains tractable for $d\le 3$ and $N_f \le {\cal O}(10^3)$ 
simply using a regular discretization of the feature space $(\n,z)\subset[-1,1]^d$ as shown in the next section. 
\begin{figure}[ht]
\includegraphics[width=\columnwidth]{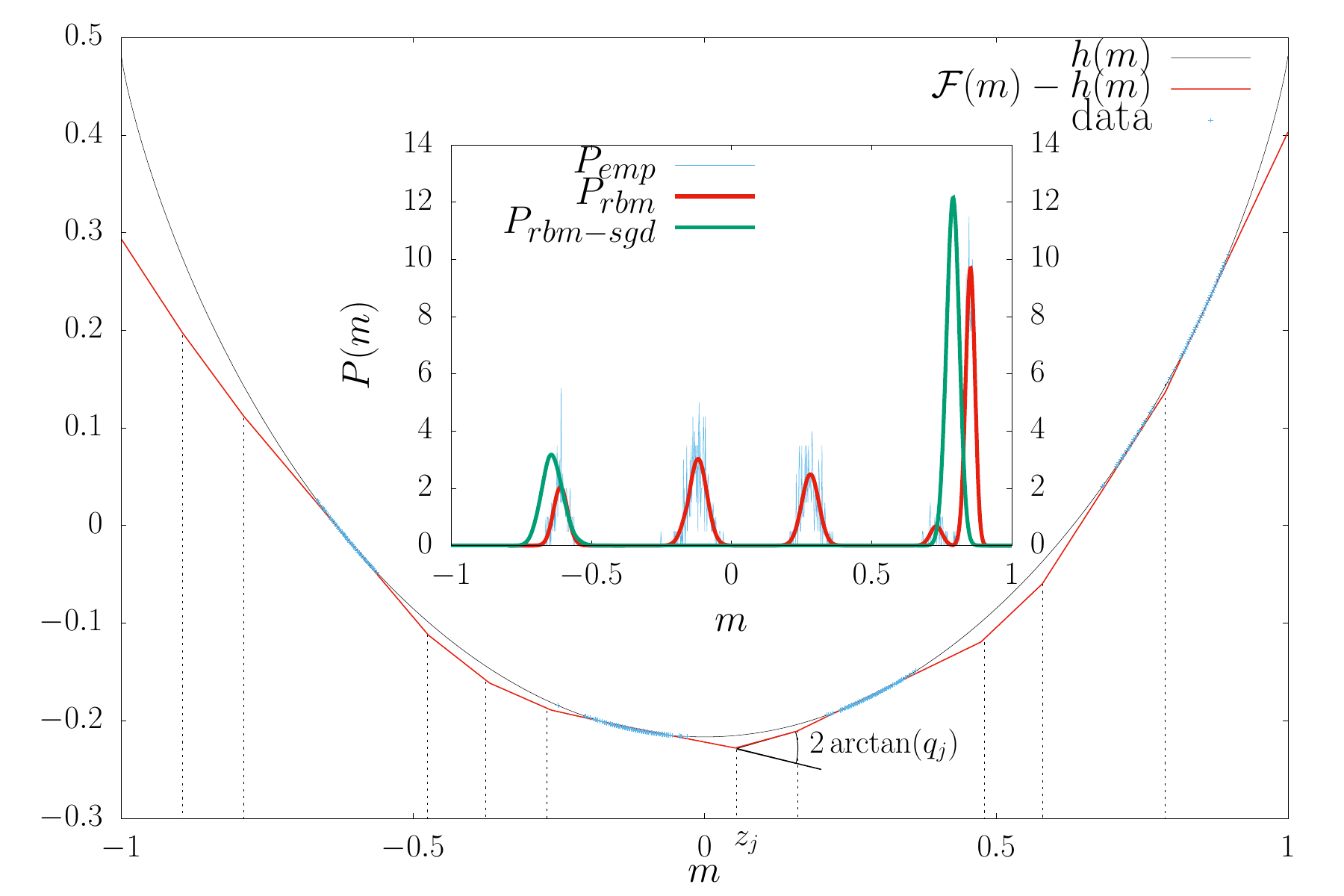}
\caption{$1$-d intrinsic data ($N_v=10^3$) with $5$ clusters solved with $N_h=20$ predefined features thanks to a natural gradient ascent of the LL.
  Dotted lines indicate location of features with nonvanishing weights $q_j$.
  The feature contributions $\F(m)-h(m)$ to the free energy are seen to regress $h(m)$ on the data.
  The resulting distribution is shown (red) on the inset with the empirical training distribution (blue) and the \emph{failed} result of a standard RBM training (green).  \label{fig:1drbm}}
\end{figure}
Additionally, an even more tractable approach bypassing the computation of $Z[\Theta]$, based on a linear regression seems plausible according to the following observations.
In terms of the Coulomb charges density
\begin{equation}\label{eq:rho-Theta}
\rho(\m\vert\Theta) = \int d\n dz\ \q(\n,z)\delta\bigl(\n^T\m-z\bigr),
\end{equation}
resulting from  a distribution $\q(\n,z)$ of uniformly charged hyperplanes
the marginal distribution of $\m$ reads
\begin{align*}
  P(\m\vert\Theta) &= \frac{1}{Z[\Theta]}e^{-N_v\F[\m\vert\Theta]},\\[0.2cm]
  &= \frac{e^{N_v\bigl(\Sc(\m)+\int d\m'\rho(\m'\vert\Theta)K_d(\vert\m-\m'\vert)-\F^\perp[\m\vert\Theta]\bigr)}}{Z[\Theta]} 
\end{align*}
with $K_d(\vert\m-\m'\vert)$ the inverse of the $d$-dimensional Laplacian $\nabla_d^2$ operator (See SM, Appendix~\ref{app:Coulomb}).
Assuming for the moment that $\rho$ is not restricted to be of the specific RBM form~(\ref{eq:rho-Theta}), this relation can be explicitly inverted to match
any smoothed version $\hat p_\epsilon(\m)$ of the empirical distribution $\hat p(\m)$:
\begin{equation}\label{eq:rhosol}
\rho(\m\vert\Theta) = \nabla_d^2\bigl(\frac{1}{N_v}\log\hat p_{\epsilon}(\m)-\Sc[\m]+\F^\perp[\m\vert\Theta]\bigr)
\end{equation}
up to surface terms,
provided that $\F^\perp$ is independent of $\rho$.
\begin{figure}[ht]
\includegraphics[width=\columnwidth]{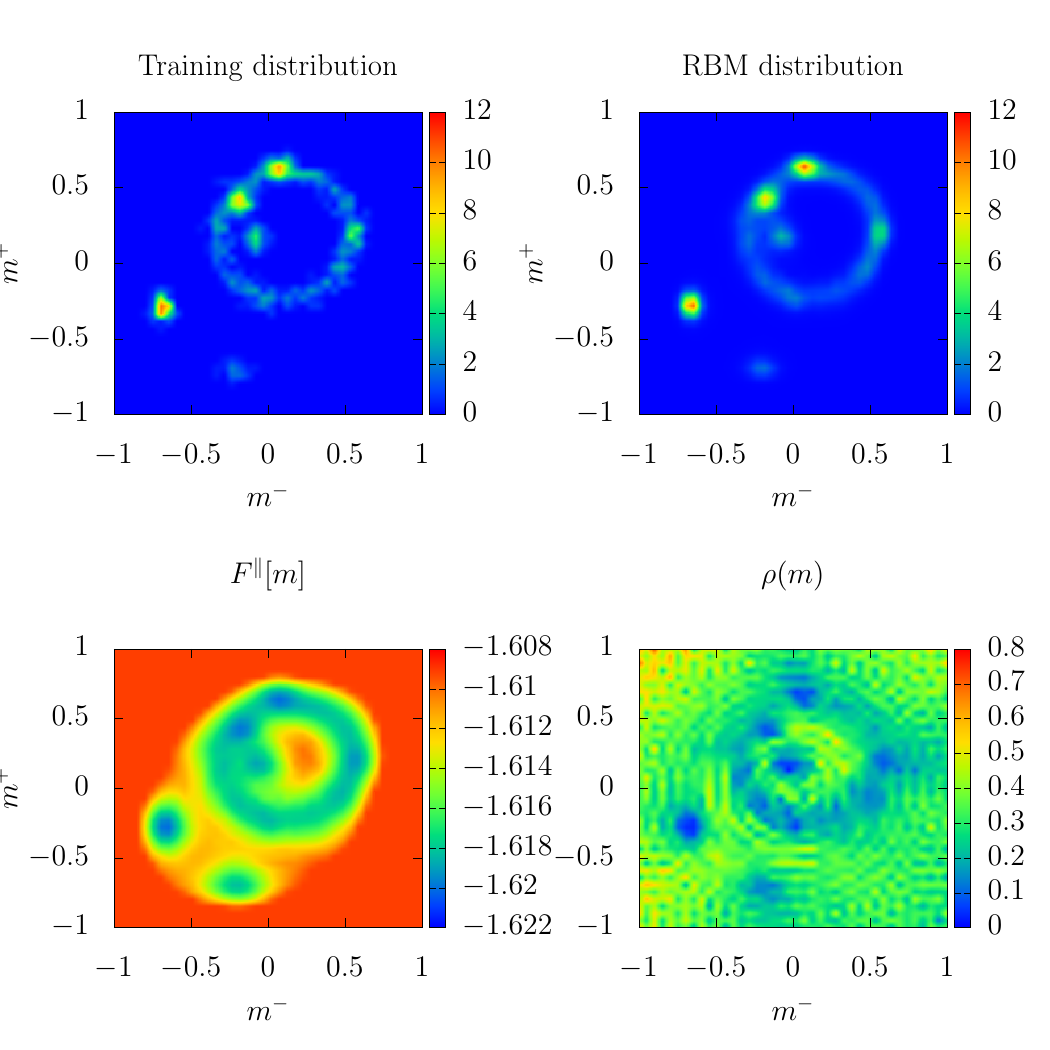}
\caption{$2$-d intrinsic dataset ($N_v=10^3$) with $6$ point-like clusters and a circular one (upper left) and corresponding RBM density (upper right) found with $N_h=900$ predefined features, along with its
   free energy landscape (bottom left) and Coulomb charges distribution (bottom right) \label{fig:2drbm}}
\end{figure}
Doing that leads to overfit the data with a density of Coulomb charges concentrated on the faces of the
Voronoi cells enclosing the data points (see SM, Appendix~\ref{app:interpol}). To be meaningful this solution has to be projected on the ``RBM" space, i.e. a density $\rho$ of the form~(\ref{eq:rho-Theta})
corresponding to a finite number of features.
The fact that any distribution $\rho$ can be approximated to arbitrary precision by such a superposition of charged hyperplanes relates to the property that the
RBM is a universal approximator~\cite{LeRoBe}.
The appropriate metric to perform such a projection is the Fisher metric~\cite{Amari} and this ends up being equivalent to minimizing the Kullback-Leibler
divergence ($D_{\rm KL}$) between $\hat p(\m)$ and $P(\m\vert\Theta)$
i.e to maximizing the LL. Nonetheless, if we expect the optimal solution to be
very close to $\hat p$, we may use directly the Fisher metric estimated at the empirical point $\hat p$ thereby turning the problem into the following linear
regression 
\begin{equation}\label{eq:regression}
\Theta^\star = \argmin_{\bq} {\mathbb E}_{\hat p}\Bigl[\bigl\vert\F^\perp[\m]-\Sc[\m] -\sum_{j=1}^{N_h}q_j V_j[\m]\bigr\vert^2\Bigr]
\end{equation}
of $\F^\perp[\m]-\Sc[\m]$ on the score variables $V_j[\m]\egaldef\frac{\partial\log(P(\m\vert\Theta))}{\partial q_j}$ conjugate to $q_j$ (see SM, Appendix~\ref{app:LR}).
\paragraph*{Study cases.}
\textemdash~To illustrate these statements first consider a dataset supported by a $1$-d subspace given by the vector $u_i  = 1/\sqrt{N_v}$ with unbiased fluctuations along other directions.
A rank one $W = w_1 u^1v^{1T}$ is assumed since we expect transverse modes to vanish from the linear stability analysis of the training given in~\cite{DeFiFu2018}. 
The relation~(\ref{eq:malpha}) reduces then to the magnetization $m = \tanh(\mu)$ along $u$
leading in the Coulomb formulation to
\[
  \F[m\vert\bq] = h(m)  - \sum_{j=0}^{N_h} q_j \vert m-z_j\vert,\qquad (q_j\ge 0)
\]
with $h(m)=\frac{1}{2}(1\pm m)\log(1\pm m)$. The natural gradient ascent of the LL yields optimal solution as the one shown on Fig.~\ref{fig:1drbm}.
As is manifest on Fig.~\ref{fig:1drbm} the result is the linear regression~(\ref{eq:regression}) of $\F^\perp[\m]-\Sc[\m] = h(m)$ in terms of a piecewise linear function,
where the break points correspond to the locations $z_j$ 
of the relevant features and $q_j$ the corresponding break of slope at these points. This involves, however, an implicit regularization which will be studied elsewhere,
in order to maintain the regions free of data below $h(m)$ in order to stay away from first-order transitions where the local Fisher metric  would cease to be a meaningful approximation to the $D_{\rm KL}$.
%\subsection{Intrinsic $2$-d case}
As a $2$-d example we consider data concentrated in the subspace spanned by the vectors $u_i^1 = 1/\sqrt{N_v}$ and $u_i^2 = (-1)^i/\sqrt{N_v}$ with irrelevant transverse fluctuations,
hence assuming now $W = w_1 u^1v^{1T}+w_2 u^2v^{2T}$. We have then a finite magnetization $(m_1,m_2)$ along each direction and
the free energy considered in the Coulomb formulation reads  
\begin{align*}
  \F[\m\vert\Theta] &= \frac{1}{2}[h(m^+)+h(m^-)]\\
  &- \sum_{j=1}^{N_h} q_j\vert m_1\cos(\omega_j)+m_2\sin(\omega_j)-z_j\vert,
\end{align*}
where $m^{\pm}=m_1\pm m_2\in[-1,1]$ and  $\omega_j\in[0,\pi[$ are the angles made by the charged lines with the $m_2$ axis.
The result of the natural gradient ascent of the LL  
is shown on Fig.~\ref{fig:2drbm}. Here a large number of features $(\omega_j,z_j)\in[0,\pi]\times[-1,1]$ have been predefined on a regular lattice
in order to obtain a continuous charge distribution and a smooth free energy landscape (see more details in SM, Appendix~\ref{app:exp}). 
Finally, in both study cases the standard RBM training fails for two distinct reasons unveiled by the Coulomb picture (see SM, Appendix~\ref{app:exp}):
(i) the Gibbs sampling is plagued by the presence of first-order phase transitions with respect to an annealing temperature;
(ii) the charged hyperplanes get easily trapped by Coulomb barriers formed by the clusters of data , a pitfall bypassed by the convex ``Coulomb'' relaxation.

\paragraph*{Discussion.} \textemdash~The physical picture of the RBM emerging here, in addition to identifying and
disentangling via Eqs.~(\ref{eq:nuj},\ref{eq:nj}) the role played by some key factors, underlines the importance
of two distinct aspects of learning a high dimensional distribution : the ordered part corresponding
to global statistical patterns and the fluctuations around these patterns encoding possibly short range correlations or corresponding to noise.
Under a flat intrinsic space hypothesis our formalism decouples them and gives indications of how to learn them separately in order to obtain high quality models that are needed in scientific applications,
when default RBM algorithms are thwarted by low dimensional global patterns as we see in our experiments. Among many possible developments we foresee
that the ``Coulomb'' convex relaxation could be used to fine-tune some otherwise poorly trained RBM, and opens the intriguing 
possibility of tackling unsupervised learning via regularized linear regressions.

\paragraph*{Acknowledgments} 
 A.D. was supported by the Comunidad de Madrid and the Complutense University of Madrid (Spain) through the Atracción de Talento program (Ref. 2019-T1/TIC-13298).

\bibliographystyle{unsrt}
\bibliography{rbm}

\begin{appendix}
\section{Canonical ensemble with magnetization constraints}\label{app:Fpara}
The decomposition of the vector of visible variables $\bs$ on the left singular basis
\begin{equation}\label{def:projs}
s_\alpha \egaldef \frac{1}{\sqrt{N_v}}\sum_{i=1}^{N_v} s_i u_i^\alpha,
\end{equation}
coincides with $m_\alpha$ for $\alpha=1,\ldots d$ by definition of the magnetization constraints.
We look for a change of variables 
$\bs\longrightarrow (\m,\bs^\perp)$ where the original spin variables are replaced by the set 
of $\m = \{m_\alpha,\alpha=1,\ldots d\}$
and ${\cal N}[\m]$ transverse spin variables.
Let us denote by ${\mathbb E}_{\bm{s}\sim {\mathbb U}}$ the expectation taken when the original spin variables are iid, $s_i\sim {\mathbb U}_{\{-1,1\}}$.
The change of measure is made by looking at the prior distribution over the original spin variables:  
\[
P_{\rm{prior}}[\bs] = \frac{1}{2^{N_v}} = P_{\rm{prior}}[\bs^\perp\vert \m]P_{\rm{prior}}[\m],
\]
where 
\begin{equation}\label{eq:prior}
P_{\rm{prior}}[\m] = {\mathbb E}_{\bm{s}\sim {\mathbb U}}\Bigl[\prod_{\alpha=1}^d \delta(s_\alpha - m_\alpha)\Bigr] = e^{N_v({\cal S}[\m]-\log 2)}
\end{equation}
represents the density of states (normalized to one) associated with the magnetization constraints $\m$, ${\cal S}[\m]$ 
the configuration entropy associated with these magnetizations and 
\[
P_{\rm{prior}}[\bs^\perp\vert\m] \egaldef \frac{1}{2^{\N[\m]}},
\]
with ${\cal N}[\m]  = N_v S[\m]/\log(2)$ representing
the remaining number of degrees of freedom $\bs^\perp$ taken out of the $N_v$ initial ones.
Note that there is a formal difficulty here because the size of the transverse variables vector  $\bs^\perp$ depends explicitly on $\m$.
This however is not really a problem if 
consider $\bs^\perp$ to be a vector of size $N_v$ where the last $N_v-\N(\m)$ bits are frozen arbitrarily to $1$, which is done in practice by defining the prior distribution
\[
P_{\rm{prior}}[\bs^\perp\vert\m] \egaldef \frac{1}{2^{\N[\m]}}\prod_{\ell=\N[\m]+1}^{N_v} \delta(s_\ell^\perp-1).
\]
To avoid additional burden on the notations we keep this as implicit and $\bs^\perp$ always refers to the set of non-frozen variables.
We want here to determine $\Sc[\m]$ from~(\ref{eq:prior}).
We have
\[
{\mathbb E}_{\bm{s}\sim {\mathbb U}}[s_\alpha]=0\qquad\text{and}\qquad{\mathbb E}_{\bm{s}\sim {\mathbb U}}[s_\alpha s_\beta] = \frac{1}{N_v}\delta_{\alpha\beta},
\]
the second relation resulting from the orthogonality of the $\bu^\alpha$ vectors.
As a result, for large $N_v$ we have
\begin{equation}\label{eq:fgm}
P_{\rm{prior}}[\m] = \frac{1}{(2\pi)^{d/2}}\exp\Bigl(-\frac{N_v}{2}\sum_{\alpha=1}^d m_\alpha^2\Bigr).
\end{equation}
This is valid as long as the magnetization are not too large ($m_\alpha={\cal O}(1/\sqrt{N_v})$). To study the regime where modes condense, 
i.e. when $m_\alpha={\cal O}(1)$, we have to resort to large deviations estimations~\cite{Touchette}.
With $d$ assumed to be ${\cal O}(1)$, as $N_v\to\infty$ we expect in this regime a behaviour of the form   
\[
P_{\rm{prior}}[\m] \asymp e^{-N_v {\cal I}[\m]},
\]
where ${\cal I}[\m]$ called the rate function, has $0$ as minimum value and can be determined in the present situation thanks to the G\"artner-Ellis 
theorem from the moment generating function of $P_{\rm{prior}}[\m]$.
Denoting by $\bm{\mu} = {\cal O}(1)$ a conjugate $d$-dimensional vector and assuming that we can make sense of the following limit
\begin{align*}
\Phi[\bm{\mu}] &\egaldef \lim_{N_v\to\infty}\frac{1}{N_v}\log\left({\mathbb E}_{\bm{s}\sim {\mathbb U}}\Bigl[ e^{ N_v\sum_{\alpha=1}^d m_\alpha(\bm{s})\mu_\alpha} \Bigr]\right),\\[0.2cm]
&= \lim_{N_v\to\infty}\frac{1}{N_v}\sum_i\log\cosh\Bigl(\sqrt{N_v}\sum_{\alpha=1}^d u_i^\alpha\mu_\alpha\Bigr),
\end{align*}
${\cal I}[\m]$ is then simply given by the Legendre-Fenchel transform of $\Phi$:
\[
{\cal I}[\m] = \bm{m}\bm{\mu}[\m]^T - \Phi\bigl[\bm{\mu}[\m]\bigr],
\]
with $\bm{\mu}[\m]$ implicitly given by (in principle when $N_v\to\infty$)
\begin{align}
m_\alpha &= \lim_{N_v\to\infty}\frac{1}{\sqrt{N_v}}\sum_{i=1}^{N_v} u_i^\alpha\tanh\Bigl(\sqrt{N_v}\sum_{\beta=1}^d u_i^\beta \mu_\beta\bigr),\nonumber\\[0.2cm]
&= {\mathbb E}_{{\bm u}\sim p_{\bm u}}\Bigl[u^\alpha\tanh\Bigl(\sum_{\beta=1}^d u^\beta \mu_\beta\bigr)\Bigr] \label{eq:mu}
\end{align}
where we assume in the last equality some limit $p_{\bm u}$ of the joint empirical distribution of $u^\alpha= \sqrt{N_v}u_i^\alpha$ when $N_v\to\infty$.
From the small $\m$ behaviour given in~(\ref{eq:fgm}) we finally have determined the configuration entropy as
\begin{align*}
\Sc[\m] &= -{\cal I}[\m] +\log(2)+\frac{d}{2N_v}\log(2\pi),\\[0.2cm]
&= \Phi[\bm{\mu}[\m]] - \m^T \bm{\mu}[\m]+\log(2)+{\cal O}\Bigl(\frac{1}{N_v}\Bigr).
\end{align*}
Note that practically speaking  we use  finite $N_v$ estimates of $\Phi$ and $m_\alpha$
so that the preceding relation is in fact valid up to some ${\cal O}(1/\sqrt{N_v})$ corrections w.r.t. limit defined 
by some hypothetical $p_{\bm u}$ when $N_v\to\infty$.

\section{Longitudinal and transverse free energy}\label{app:FLT}
In order to disentangle the contributions of the collective modes materialized by some magnetization $\m$ along some directions $\bu^\alpha, \alpha=1,\ldots d$
from the noise corresponding to the fluctuations of the transverse variables $\bs^\perp$ 
we assume first to be able to rewrite the Hamiltonian corresponding to the visible distribution~(\ref{eq:Prbm})
\[
P[\bs\vert\Theta] = \frac{e^{-\Hc[\bs\vert\Theta]}}{Z[\Theta]},
\]
in terms of the new degrees of freedoms
\[
\Hc[\bs\vert\Theta] = \Hc[\m,\bs^\perp\vert\Theta],
\]
such that the joint distribution takes the form
\[
P[\m,\bs^\perp\vert\Theta] = \frac{e^{-\Hc[\m,\bs^\perp\vert\Theta]}}{Z[\Theta]}.
\]
This in turn can be written 
\begin{equation}\label{eq:pmbs}
P[\m,\bs^\perp\vert\Theta] = P[\bs^\perp\vert\m,\Theta]P[\m\vert\Theta],
\end{equation}
with
\[
  P[\m\vert\Theta] = \sum_\bs P[\bs]\prod_{\alpha=1}^d\delta(s_\alpha-m_\alpha) 
  \egaldef \frac{e^{-N_v\F[\m\vert\Theta]}}{Z[\Theta]},
\]
after introducing the canonical free energy $\F[\m\vert\Theta]$.  
Let us denote by 
\begin{equation}\label{eq:Hcond}
\Hc[\bs^\perp\vert\m,\Theta] \egaldef \Hc[\m,\bs^\perp\vert\Theta]-\Hc_0[\m\vert\Theta],
\end{equation}
the ``conditional" Hamiltonian, where $\Hc_0[\m\vert\Theta]$ is at this point an arbitrary function of $\m$ independent of $\bs^\perp$.
We chose it as to contain only contributions from the longitudinal magnetization, i.e. coincides with the constant part of $\Hc[\m,\bs^\perp\vert\Theta]$ w.r.t. $\bs^\perp$
when neglecting the singular values $w_\beta$ of $W$ and the ${\mathcal O}\Bigl(1/\sqrt{N_v}\Bigr)$ residual transverse magnetizations $m_\beta$ for $\beta>d$ resulting from the constraints (see next Section).
Rewriting the Hamiltonian in terms of the SVD components $s_\alpha$ and $\eta_\alpha$ respectively of the visible variables and biases
\begin{align*}
  &\Hc[\bs\vert\Theta] = \sum_{i=1}^{N_v}\eta_i s_i - \sum_{j=1}^{N_h}\log\cosh\Bigl(\sum_{i=1}^{N_v}W_{ij}s_i -\theta_j\Bigr)\\[0.2cm]
  &= N_v\sum_{\alpha=1}^{N_v}\eta_\alpha s_\alpha - \sum_{j=1}^{N_h}\log\cosh\Bigl(\sqrt{N_v}\sum_{\alpha=1}^{N_v}w_\alpha s_\alpha v_j^\alpha -\theta_j\Bigr),
\end{align*}
this leads to the definition
\begin{equation}\label{eq:H0}
\Hc_0[\m\vert\Theta] = N_v\bigl(\sum_{\alpha=1}^d\eta_\alpha m_\alpha -V(\m\vert\Theta)\bigr),
\end{equation}
with
\[
V(\m\vert\Theta)\bigr)\egaldef \frac{1}{N_v}\sum_{j=1}^{N_h}\log\cosh\Bigl(\sqrt{N_v}\sum_{\alpha=1}^d w_\alpha m_\alpha v_j^\alpha -\theta_j\Bigr).
\]
In terms  of $\Hc_0[\m\vert\Theta]$ the free energy reads
\begin{align*}
  \F[\m\vert\Theta] &= \frac{1}{N_v}\Bigl[\Hc_0[\m\vert\Theta]-\log\Bigl(\sum_{\bs^\perp}e^{-\Hc[\bs^\perp\vert\m,\Theta]}\Bigr)\Bigr],\\[0.2cm]
  &= \frac{1}{N_v}\Bigl[\Hc_0[\m\vert\Theta]-\Sc[\m]-\log\Bigl(\frac{\sum_{\bs^\perp}e^{-\Hc[\bs^\perp\vert\m,\Theta]}}{2^{\N[\m]}}\Bigr)\Bigr]\\[0.2cm]
  &= \F^\parallel[\m\vert\Theta]+\F^\perp[\m\vert\Theta]
\end{align*}
where we have introduced respectively the longitudinal and transverse free energy:
\begin{align*}
  \F^\parallel[\m\vert\Theta] &\egaldef \frac{1}{N_v}(\Hc_0[\m\vert\Theta]-\Sc[\m]),\\[0.2cm]
  \F^\perp[\m\vert\Theta] &\egaldef -\frac{1}{N_v}\log\Bigl(\frac{1}{2^{\N[\m]}}\sum_{\bs^\perp} e^{-\Hc[\bs^\perp\vert \m,\Theta]}\Bigr).
\end{align*}
The longitudinal free energy of the system is the free energy of the system for a given magnetization $\m$
when the transverse magnetization and interactions among the $s_\ell^\perp$ are neglected. These indeed are expected to
be small by definition of  the intrinsic space. Vanishing interactions corresponds to $\Hc\bigl[\bs^\perp\vert \m,\Theta\bigr]=0$, in which case $\F^\parallel[\m\vert\Theta]$
coincides with $\F[\m\vert\Theta]$. 
Non-vanishing interactions are accounted for by the transverse free energy.
%The distribution of the transverse variables conditionally to the longitudinal ones are then given by 
%\[
%P[\bs^\perp\vert\m,\Theta] = \frac{e^{-\Hc[\bs^\perp\vert \m,\Theta]}}{Z^\perp[\m,\Theta]}P_{\rm{prior}}[\bs^{\perp}\vert\m],
%\]
%with the transverse partition function
%\[
%Z^\perp[\m,\Theta] = \sum_{\bs^\perp}e^{-\Hc[\bs^\perp\vert \m,\Theta]}P_{\rm{prior}}[\bs^{\perp}\vert\m].
%\]
%Accordingly we define the transverse free energy as
%\begin{align*}
%  \F^\perp[\m\vert\Theta] &\egaldef -\frac{1}{N_v}\log\bigl(Z^\perp[\m,\Theta]\bigr),\\[0.2cm]
%  &= -\frac{1}{N_v}\log\Bigl(\frac{1}{2^{\N[\m]}}\sum_{\bs^\perp} e^{-\Hc[\bs^\perp\vert \m,\Theta]}\Bigr),\\[0.2cm]
%  &= \frac{1}{N_v}\Bigl(\Sc[\m]-\log\bigl(\sum_{\bs^\perp} e^{-\Hc[\bs^\perp\vert \m,\Theta]}\bigr)\Bigr)
%\end{align*}
%which is consistent with~(\ref{eq:pmbs2}) and the decomposition  
%\[
%\F[\m\vert\Theta] = \F^\parallel[\m\vert\Theta]+\F^\perp[\m\vert\Theta]
%\]
%holds.

\section{Effective Hamiltonian}\label{app:EffH}
To enter further into the description of transverse fluctuations we need to specify the transverse degrees of freedom $\bs^\perp$ and  the
way they interact through $\Hc[\bs^\perp\vert\m,\Theta]$ in the form  of an effective Hamiltonian.
For $\beta>d$ the components $\bs_\beta$ given in~(\ref{def:projs}) are stochastic variables and we denote them by
$s_\beta[\bs^\perp\vert\m]$, i.e. a mapping to be defined of  transverse variables to transverse projections given some  magnetization $\m$.
This mapping cannot be determined exactly but since we look for an effective theory we consider 
a linear map i.e. of the form
\begin{equation}\label{eq:sperp}
s_\beta[\bs^\perp\vert\m] = m_\beta[\bm{\mu}]+\sum_{\ell=1}^{\N[\m]} c_\beta^\ell s_\ell^\perp,\qquad \beta=d+1,\ldots N_v,
\end{equation}
where the prior of these new variables is to be iid with $s_\ell^\perp\sim {\mathbb U}_{\{-1,1\}}$.
Under the magnetization constraints \emph{only} we have (see Section~\ref{app:Fpara})
\begin{align*}
{\mathbb E}[s_i\vert\m] &= m_i[\bm{\mu}] \\[0.2cm] 
\Cov(s_i,s_j\vert\m) &= (1-m_i^2[\bm{\mu}])\delta_{ij}+{\mathcal O}\Bigl(\frac{1}{N_v}\Bigr)
\end{align*}
with
\[
m_i[\bm{\mu}] = \tanh\Bigl(\sqrt{N_v}\sum_{\alpha=1}^d \mu_\alpha u_i^\alpha\Bigr).
\]
As a result  
$m_\beta[\bm{\mu}]$ (for $\beta>d$) represents the prior bias resulting from the magnetizations constraints~(\ref{eq:mu}) 
\begin{equation}\label{eq:mbeta}
m_\beta[\bm{\mu}] = \frac{1}{\sqrt{N_v}}\sum_{i=1}^{N_v} u_i^\beta\tanh\Bigl(\sqrt{N_v}\sum_{\alpha=1}^d \mu_\alpha u_i^\alpha\Bigr),\ \forall\beta>d  
\end{equation}
as a function of $\{\mu_\alpha,\alpha=1,\ldots d\}$ solution to equation~(\ref{eq:mu}) and 
are ${\cal O}(1/\sqrt{N_v})$.
The second set of constraints comes from the set of prior covariances between the $s_\beta[\bs^\perp\vert\m]$ for $\beta>d$ that have to be maintained to properly account for
the transverse degrees of freedom.
These are diagonal at leading order:
\begin{align}
  \Cov(s_\beta[\bs^\perp\vert\m],s_\gamma[\bs^\perp\vert\m]) &= \frac{1}{N_v}\sum_{i=1}^{N_v} {u_i^{\beta}}^2 (1-m_i^2[\bm{\mu}])\delta_{\beta\gamma}\nonumber\\[0.2cm]
  &+{\mathcal O}\Bigl(\frac{1}{N_v^2}\Bigr)\label{eq:cs2}.
\end{align}
A simple way to maintain at best these covariances in the new representation is to associate a binary variable $s_\ell^\perp$ with the first $\N[\m]$ principal axes of the previous covariance matrix
neglecting the remainder. Since the covariance resulting from~(\ref{eq:sperp})
reads
\[
  \Cov(s_\beta[\bs^\perp\vert\m],s_\gamma[\bs^\perp\vert\m]) = \sum_{\ell=1}^{\N[\m]} c_\beta^\ell c_\gamma^\ell,
\]
the vector $\bm{c}_\ell$ can be chosen as a principal axes of the covariance matrix~(\ref{eq:cs2}) normalized to the standard deviation along the same axes which are ${\mathcal O}\Bigl(1/\sqrt{N_v}\Bigr)$
so that $c_\ell^\beta$ coefficients are ${\mathcal O}\Bigl(1/N_v\Bigr)$.
Now that the transverse variables are unambiguously defined we can obtained their effective Hamiltonian by expanding $\Hc[\bs^\perp\vert\m]$ defined in~(\ref{eq:Hcond})
at second order in $s^\perp$. From the definition~(\ref{eq:Hcond},\ref{eq:H0}) there is a zero order term $\F_0^\perp[\m\vert\Theta] = {\mathcal O}(1/\sqrt{N_v})$ with convoluted expression
which we give only the first terms, contributing to the transverse free energy and the first
and second order terms corresponds to a conventional Hamiltonian of a disordered Ising model:
\begin{align*}
  \Hc[\bs^\perp\vert\m,\Theta] &\approx \Hc_{\rm eff}\bigl[\bs^\perp\vert \m,\Theta\bigr]\\[0.2cm]
  &= N_v\F_0^\perp[\m\vert\Theta]\\[0.2cm]
  &+\sum_{\ell=1}^{\cal N[\m]}
\eta_\ell^\perp[\m,\Theta]s_\ell^\perp+\sum_{\ell,\ell'=1}^{{\cal N}[\m]} W^\perp_{\ell\ell'}[\m,\Theta] s_\ell^\perp  s_{\ell'}^\perp.
\end{align*}
with
\begin{align*}
\F_0^\perp[\m\vert\Theta] &= \sum_{\beta=d+1}^{N_v} (\eta_\beta+w_\beta\bar m_\beta)m_\beta[\mu]\\[0.2cm]  
\eta_\ell^\perp[\m,\Theta] &= N_v\sum_{\beta=d+1}^{N_v} c_\beta^\ell(\eta_\beta -w_\beta\bar m_\beta) \\[0.2cm]
W_{\ell,\ell'}^\perp[\m,\Theta] &= -N_v\sum_{j=1}^{N_h}(1-\bar m_j^2)\sum_{\beta,\gamma=d+1}^{N_v}w_\beta w_\gamma c_\beta^\ell c_\gamma^{\ell'}v_j^\beta v_j^\gamma,
\end{align*}
after introducing the notations:
\begin{align*}
\bar m_j &= \tanh\Bigl(\sqrt{N_v}\bigl(\sum_{\alpha=1}^{d} w_\alpha m_\alpha v_j^\alpha+\sum_{\beta=d+1}^{N_v} w_\beta m_\beta[\bm{\mu}] v_j^\beta\bigr)-\theta_j\Bigr),\\[0.2cm]
\bar m_\beta &= \frac{1}{\sqrt{N_v}}\sum_{j=1}^{N_h} \bar m_j v_j^\beta. 
\end{align*}
$\eta_\ell^\perp[\m,\Theta]$ is potentially ${\cal O}(1)$ while $W_{\ell,\ell'}^\perp[\m,\Theta]$ is ${\cal O}\Big(\frac{1}{\sqrt{N_v}}\Bigr)$.

To make connection with data, i.e. given a configuration  $\bs$ with magnetization $m_\alpha,\alpha=1,\ldots d$, from which are extracted the transverse magnetization
$\{m_\beta[\bm{\mu}],\beta=d+1,\ldots N_v$ by solving the equations~(\ref{eq:mu}), the $\bs^\perp$ are constructed as follows. First let
for each $\ell=1,\ldots,{\cal N}[\m]$
\[
m_\ell^\perp[\bs] =  \sum_{\beta=d+1}^{N_v} (s_\beta-m_\beta[\bm{\mu}]) u_\beta^{\perp\ell}
\]
(thresholded to $1$ [resp. $-1$] when bigger than $1$ [resp. lower than $-1$])
the magnetization of the configuration $\bs$ along this mode. This allows us to define the probability
\[
p_\ell[\bs] = \frac{1+m_\ell^\perp[\bs]}{2}.
\]
Then 
\[
s_\ell^\perp = 2\tau_\ell-1,\qquad\forall\ell=1,\ldots {\cal N}[\m]
\]
with $\tau_\ell$ a Bernoulli variable of parameter $p_\ell[\bs]$, gives us a set of spin variables fulfilling our needs. 

\section{Coulomb interaction picture}\label{app:Coulomb}
Let us consider the green function for the 
$d$-dimensional Laplacian $\nabla_d^2$
\[
K_d(\vert \m-\m'\vert) =
\begin{cases}
  \DD \frac{1}{2}\vert m-m'\vert,\qquad\qquad\qquad\ (d=1)\\[0.2cm]
  \DD \frac{1}{2\pi}\log\vert\m-\m'\vert,\qquad\qquad (d=2)\\[0.2cm]
  \DD -\frac{\Gamma\bigl(\frac{d}{2}-1\bigr)}{4\pi^{d/2}\vert \m-\m'\vert^{d-2}},\qquad (d>2)
  \end{cases}
\]
which by definition is solution of 
\[
\nabla_d^2 K_d(\vert \m-\m'\vert) = \delta(\m-\m').
\]
We can use it to rewrite $V[\m\vert\Theta]$ up to an irrelevant constant term as
\begin{equation}\label{eq:Vmrho}
V[\m\vert\Theta] = \int d^d\m' \rho(\m'\vert\Theta) K_d(\vert \m-\m'\vert)
\end{equation}
where $\rho$ is the source term of the Poisson equation
\[
\nabla_d^2 V[\m\vert\Theta] = \rho(\m)
\]
hence giving a density of Coulomb charges  
\begin{align*}
  \rho(\m\vert\Theta) = \sum_{j=1,\alpha=1}^{N_h,d}&w_\alpha^2{v_j^\alpha}^2\Bigl(1\\
  &-\tanh^2\bigl(\sqrt{N_v}\sum_{\beta=1}^d w_\beta m_\beta v_j^\beta -\theta_j\bigr)\Bigr).
\end{align*}
To make sense of this quantity first remark that the function
\[
\delta_\nu(x) \egaldef \frac{\nu}{2}\bigl[1-\tanh^2(\nu x)\bigr] \lra_{\nu\to\infty} \delta(x)
\]
represents a normalized $1$-d narrow density of width $\nu^{-1}$ such that $\rho$ can be expressed as
\begin{equation}\label{eq:rhom}
\rho(\m\vert\Theta) = \frac{2}{N_v}\sum_{j=1}^{N_h} \nu_j \delta_{\nu_j}\bigl(\n_j^T\m-z_j\bigr)
\end{equation}
with $\nu_j$ and $n_j$ given by equation~(\ref{eq:nuj},\ref{eq:nj}).
In this form, $\rho$ is readily a superposition of $N_h$ uniformly charged hyperplanes of finite width. Each hyperplane $j$
being defined by a normal vector $\n_j$, an offset $z_j=\theta_j/\nu_j$ from the origin, a finite width $\nu_j^{-1}$ and a (hyper)surface charge density $2\nu_j/N_v$.
Furthermore this can be decomposed into more elementary charged hyperplanes of zero width.
Equation~(\ref{eq:Vmrho}) now rewrites
\[
V[\m\vert\Theta] = \frac{2}{N_v}\sum_{j=1}^{N_h} \nu_j\int d^d\m' \delta_{\nu_j}\bigl(\n_j^T\m'-z_j\bigr) K_d(\vert \m-\m'\vert)
\]
For each term $j$, writing $\m' = (\n_j^T\m')\n_j+\m'^\perp = z\n_j+\m'^\perp$, the transverse integration of  $\m'^\perp$ yields
\begin{align*}
  &\int d\m' K_d\bigl(\vert \m-\m'\vert\bigr) =\\[0.2cm]
  &  \int dz d\m'^{\perp} K_d\bigl(\sqrt{(\n_j^T\m-z)^2+(\m^\perp-\m'^\perp)^2}\bigr) \\[0.2cm]
 &= \int dz\vert \n^T\m - z\vert. 
\end{align*}
up to an ill defined constant term after properly regularizing at large distances the integral over $\m'^\perp$. 
As a result the one particle potential takes the form
\begin{align*}
  V[\m\vert\Theta] &= \frac{2}{N_h}\sum_{j=1}^{N_h} \nu_j \int dz \delta_{\nu_j}\bigl(z-z_j\bigr) \vert\n_j^T\m- z\vert.\\[0.2cm]
  &= \int d\n dz\ \q(\n,z)\vert\n^T\m-z\vert
\end{align*}
with $\q(\n,z)$ given by equation~(\ref{eq:qntheta})
\section{Exact Coulomb charges interpolation}\label{app:interpol}
In order to interpolate exactly the empirical distribution $\hat p$ with a generalized Coulomb charges RBM based distribution it is needed to
regularize $\log(\hat p(\m))$. This can be done in many different ways.
Consider for instance 
\[
\delta_\epsilon(\m) \egaldef \frac{\exp\bigl(-\frac{\vert\m\vert^2}{2\epsilon}\bigr)}{(2\pi\epsilon)^{d/2}}
\]
with infinitesimal $\epsilon$ to approximate our point-like distribution as
\[
\hat p (\m) = \frac{1}{M}\sum_{k=1}^M\delta_\epsilon(\m-\m_k),
\]
and let
\[
q_\epsilon(k\vert \m) = \frac{\delta_\epsilon(\m-\m_k)}{\sum_l \delta_\epsilon(\m-\m_l)}.
\]
These probability weights realize a smooth partition of the space at finite $\epsilon$ with Voronoi cells ${\cal R}_k$ centered at each data point, $q_\epsilon(k\vert\m)$
representing the probability that $\m$ belongs to $k$th cell. Equipped with this notation we have
\begin{align*}
  \nabla_{\m}\delta_\epsilon(\m-\m_k) &= -\frac{\m-\m_k}{\epsilon}\delta_\epsilon(\m-\m_k)\\[0.2cm]
  \nabla_{\m}q_\epsilon(k\vert\m) &= - \frac{\m-\m_k}{\epsilon}q_\epsilon(k\vert\m)\\[0.2cm]
  &+\sum_{\ell=1}^M \frac{\m-\m_\ell}{\epsilon}q_\epsilon(k\vert\m)q_\epsilon(\ell\vert\m).
\end{align*}
As a result we get 
\[
  \nabla_d^2\log\hat p(\m) =   -\frac{1}{\epsilon}+\frac{1}{\epsilon^2}\sum_{k=1}^M \Var_{k\sim q_\epsilon(k\vert\m)}[\m_k].
\]
When $\epsilon$ becomes small compared to nearest neighbour distances this quantity becomes constant ($=-1/\epsilon$) except on the intersections between Voronoi cells, in particular on
common faces $\Vr_k\cap\Vr_\ell$
between two cells $\Vr_k$ and $\Vr_\ell$ it is
\[
\nabla_d^2\log\hat p(\m) \msim_{\epsilon\to 0} -\frac{1}{\epsilon}+\frac{\vert\m_k-\m_\ell\vert}{2\epsilon}\delta(\m\in\Vr_k\cap\Vr_\ell).
\]
Indeed, let
\[
\theta_k \egaldef \frac{1}{2}(\m_k+\m_{k+1})\qquad\text{and}\qquad \Delta_k \egaldef \frac{1}{2}(\m_{k+1}-\m_k). 
\]
For $\delta \m = \m-\theta_k$ small compared to $\Delta_k$ we have 
\[
q_k(\theta_k+\delta\m) = \frac{1}{2}\Bigl[1-\tanh\Bigl(\frac{\Delta_k^T\delta \m}{2\epsilon}\Bigr)\Bigr],
\]
leading to
\[
\Var_{k\sim q_k(\m)}[\m_k] = \vert\Delta_k\vert^2\Bigl[1-\tanh^2\Bigl(\frac{\Delta_k^T\delta\m}{2\epsilon}\bigr)\Bigr].
\]
Since $\frac{\nu}{2}\bigl[1-\tanh^2(\nu x)\bigr]$ tends to $\delta(x)$ when  $\nu\to\infty$ we arrive at the statement.
As a result the distribution of charges is composed of a constant background $+$ surface distribution on Voronoi cells intersections:
\[
\rho_{\rm bulk}(\m) = -\frac{1}{N_v\epsilon}+\frac{\vert\m_k-\m_\ell\vert}{2N_v\epsilon}\delta(\m\in\Vr_k\cap\Vr_\ell).
\]
The Voronoi cells intersecting with the boundary of the $\m$ domain induce additional surface charges which can be directly taken care of with visible bias.
Let us show how this works in $1$-d.
Let us call
\[
V(m) = \frac{1}{N_v}\log\hat p(m)
\]
which when regularized reads
\[
V(m) = -\frac{1}{N_v}\min_k \frac{(m-m_k)^2}{2\epsilon}.
\]
From what precedes, this potential can be exactly decomposed onto a set of features as
\[
V(m) = -\frac{m^2}{2N_v\epsilon}+\eta m + \sum_j q_j\vert m-z_j\vert
\]
with
\[
q_j = \frac{1}{2N_v\epsilon}(m_{j+1}-m_j),
\]
while from the limit behaviour $V'(1)$ and $V'(-1)$ of $V'(m)$ we get
\[
\eta = \frac{m_1+m_{N_v}}{2\epsilon}.
\]
\section{RBM optimization seen as a linear regression}\label{app:LR}
\begin{figure}[ht]
\centering{
\resizebox*{0.7\columnwidth}{!}{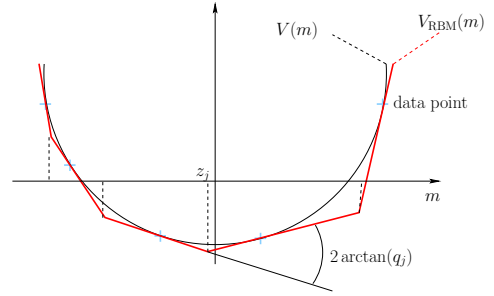}
\caption{Picture of the $V_{\rm RBM}(\m)$ (in red) at the interpolation threshold in $1$-d.\label{fig:RBM-reg}}}
\end{figure}

The projection of the empirical distribution onto the space of RBM  with finite number of features is classically done
by minimizing the Kullback Liebler divergence ($D_{\rm KL}$). If however our RBM space is chosen
with a high number of relevant features, we may expect the solution to be close enough to the empirical distribution so that a Fisher 
metric, i.e. the infinitesimal counterpart of the $D_{\rm KL}$, evaluated from  the solution or from the empirical distribution should coincide. 
In that case it might be pertinent to use it instead of the $D_{\rm KL}$. Let us formalize more precisely this projection problem.
On one hand we have the empirical measure approximated by a Coulomb based RBM model of the form 
\[
p(\m\vert\hat\rho) = \frac{1}{Z[\hat \rho]}e^{-N_v \F(\m\vert\hat\rho)}
\]
with 
\[
\F(\m\vert\hat\rho) = \F^{\perp}[\m]-\Sc[\m]-\int d\m'\hat\rho(\m')K_d(\vert\m-\m'\vert),
\]
where $\hat\rho(\m)$ is, as seen in the previous Section,
the charge density concentrated on the Voronoi cells faces coming from the empirical part $\log\hat p(\m)$ (including surface terms at the edge of the domain of $\m$). 
On the other hand we have an RBM with a pointwise distribution of features $\q(\n,z)$ yielding a free energy of the form
\[
  \F(\m\vert\Theta) = \F^{\perp}[\m]-\Sc[\m]-\int d\m'\rho(\m'\vert\Theta)K_d(\vert\m-\m'\vert),
\]
with
\[
\rho(\m\vert\Theta) = \sum_{j=1}^{N_h}q_j\delta\bigl(\n_j^T\m-z_j\bigr).
\]
Our goal is to find the (positive) weights $\{q_j,j=1,\ldots N_h\}$ such that the following distance
\begin{equation}\label{eq:inner}
D(\rho,\rho') = \int d\m_1 d\m_2 \rho(\m_1)J(\m_1,\m_2)\rho'(\m_2),
\end{equation}
between $\hat\rho$ and $\rho$ is minimized. Here the relevant metric  $J$ is the Fisher metric defined as
\begin{align}
  J\bigl[\m_1,\m_2\bigr] &= \Cov_{\m\sim p(\m\vert\Theta)}\Bigl[K_d(\vert\m-\m_1\vert),K_d(\vert\m-\m_2\vert)\Bigr],\nonumber\\[0.2cm]
  &\simeq \Cov_{\m\sim \hat p(\m)}\Bigl[K_d(\vert\m-\m_1\vert),K_d(\vert\m-\m_2\vert)\Bigr],\label{eq:empFisher}
\end{align}
approximated at the empirical point in last equation.
As we shall see this projection turns out to be a linear regression of the centered random variable
\begin{align*}
  V(\m\vert\hat\rho) &= \int d\m'\hat\rho(\m')K_d(\vert\m-\m'\vert)\\[0.2cm]
  &- \EE_{\m\sim \hat p(\m)}\Bigl[\int d\m' \hat\rho(\m')K_d(\vert\m-\m'\vert)\Bigr]
\end{align*}
onto the set of centered random variables (the score variables associated with $\bq$)
\begin{align*}
V_j(\m) &\egaldef \int d\m' \delta\bigl(\n_j^T\m'-z_j\bigr)K_d(\vert\m-\m'\vert)\\[0.2cm]
  &- \EE_{\m\sim \hat p(\m)}\Bigl[\int d\m' \delta\bigl(\n_j^T\m'-z_j\bigr)K_d(\vert\m-\m'\vert)\Bigr]    \\[0.2cm]
&= \vert \n_j^T\m-z_j\vert-\EE_{\m\sim \hat p(\m)}\Bigl[\vert \n_j^T\m-z_j\vert\Bigr].
\end{align*}
$\EE_{\m\sim \hat p(\m)}$ and $\Cov_{\m\sim \hat p(\m)}$ denote respectively empirical expectation and covariance, according to our assumption that the solution is close to $\hat p$. 
Indeed, from elementary linear algebra, the orthogonal projection $V^\parallel$ of a given  vector $\hat V$, onto a subspace spanned by a set of independent vectors $V_k$
is given by
\begin{equation}\label{eq:Vpara}
V^\parallel = \sum_{k,l} \Bigl[G^{-1}]_{kl} (V_l,\hat V) V_k 
\end{equation}
with $G_{kl} = (V_k,V_l)$ the Gram matrix of the set of vector $V_k$ for some given inner product $(\cdot,\cdot)$. Specified to our problem,
the vectors are the densities $\rho$, or equivalently the random variables $V(\m\vert\rho)$ with inner product~(\ref{eq:inner},\ref{eq:empFisher})
resulting in
\[
\bigl(V(\m\vert\rho),V(\m\vert\rho')\bigr) = \Cov_{\m\sim \hat p(\m)}\Bigl[V(\m\vert\rho),V(\m\vert\rho')\Bigr].
\]
The projection of $\hat V[\m] = V(\m\vert\hat\rho)$ is then given by $V^\parallel$ in~(\ref{eq:Vpara}) with $G$ the empirical covariance matrix of $\{V_1,\ldots V_{N_h}\}$
and $(V_k, \hat V)$ the empirical covariance between $V_k$ and $\hat V$ (if the set $V_k$ is not independent the pseudo-inverse of $G$ is taken instead of $G^{-1}$).
At this point this regression seems intractable since $\hat V$ involves a very complicated density of charge $\hat \rho$. This is not the case because
by construction we have 
\[
\int d\m' \hat\rho(\m') K_d\bigl(\vert\m'-\m\vert\bigr) = \frac{1}{N_v}\log\hat p(\m)-\Sc[\m]+\F^\perp[\m],
\] 
and $\log\hat p(\m) = \log\frac{1}{M}$ when evaluated on the data. This means that the solution to our projection problem is obtained by performing the previous linear regression with
\[
\hat V(\m) = \F^\perp[\m]-\Sc[\m] - \EE_{\m\sim \hat p(\m)}\Bigl[\F^\perp[\m]-\Sc[\m]\Bigr],
\]
so that the RBM distribution will be finally of the form
\[
P_{\rm RBM}(\m) = \frac{1}{Z}e^{-N_v\F(\m\vert\Theta)} 
\]
with
\begin{align*}
  \F(\m\vert\Theta) &= \F^\perp[\m]-\Sc[\m] - \sum_{j=1}^{N_h}q_j \vert \n_j^T\m-z_j\vert,\\[0.2cm]
  &\egaldef V(\m) - V_{\rm RBM}(\m),
\end{align*}
i.e. a difference between 2 convex potential whenever $\F^\perp[\m]$ is convex or negligible.
The interpolation point corresponding to the situation where there is a sufficient amount of Coulomb features to model exactly the empirical distribution is shown on Fig.~\ref{fig:RBM-reg}.  
In this appealing picture there is however a shortcoming.
The fact that we impose the features weights to be non-negative insures the regression curve to be convex but do not prevent it to pass above the fitted potential
in empty regions of data. The reason for this, while the information theory argument provides us with a strong  guarantee at first sight,  
is that the empirical Fisher metric is not relevant everywhere on the embedding functional space defined by the RBM features used to approximate $V(\m)$, but only on regions supported with data.
Other directions are represented by random variables which are decorrelated from the data, so the Fisher metric is not covered by~(\ref{eq:inner}) along these directions. This requires the linear regression to be complemented with some additional regularization
which as shown on Fig.~\ref{fig:RBM-reg} should involve also the distance between the derivatives of the two profiles measured at the sample points.
\section{Experiments}\label{app:exp}
The transverse free energy being absent and the entropy term independent of $\Theta\equiv\bq$, the loss optimized in the experiments is given by
\[
\L[\bq] = -{\mathbb E}_{\m\sim\hat p}\bigl[V[\m\vert\bq]\bigr] - \log\bigl(Z[\bq]\bigr),
\]
corresponding to
\[
p(\m\vert\bq) = \frac{1}{Z[\bq]}e^{-N_v\F[\m\vert\bq]}
\]
with
\[
\F[\m\vert\bq] = -\Sc[\m]-\sum_{j=0}^{N_h} q_j\vert\n_j^T\m-z_j\vert
\]
\begin{table*}[ht]
  \caption{\label{tab:LLscores} LL obtained for the RBM in the study cases}
\centering
  \begin{ruledtabular}
    \begin{tabular}{lccccccc} & $LL_{\rm{Ref}}$ & $LL_{\rm{Coulomb}}$ & $LL_{\rm{RBM}}$ &  $N_{\rm{features}}$ & $N_{\rm{epochs}}$ & $N_{pts}^d$ & $\gamma$\\
      $d=1$ &-471.45 & -479.85 & -479.82 & 5  & 5000  & 100 & 0.0001\\
            &-471.45 & -471.66 & -471.69 & 10  & 7000  & 100 & 0.0001\\
            &-471.45 & -471.48 & -471.48 & 20  & 10000  & 100 & 0.001\\
            &-471.45 & --- & -535.00 & 20 & 2500 & --- & 0.001\\
      $d=2$ &-621.90 & -623.74 & -623.61 & 49  & 113000 & 900 & 0.0001\\
            &-621.90 & -622.16 & -622.24 & 169 & 155000 & 900 & 0.0001\\
            &-621.90 & -621.97 & -631.79 & 900 & 410000 & 1600 & 0.0001\\
    \end{tabular}
  \end{ruledtabular}
\end{table*}
In the $1$-d study case, we have $n_j=1$ and the discretization concerns only $z_j = -1+2j/N_h,j=1,\ldots N_h$.
The features $j=0$ and $j=N_h$ correspond to the visible bias $\eta$.
\begin{figure}[ht]
\includegraphics[width=\columnwidth]{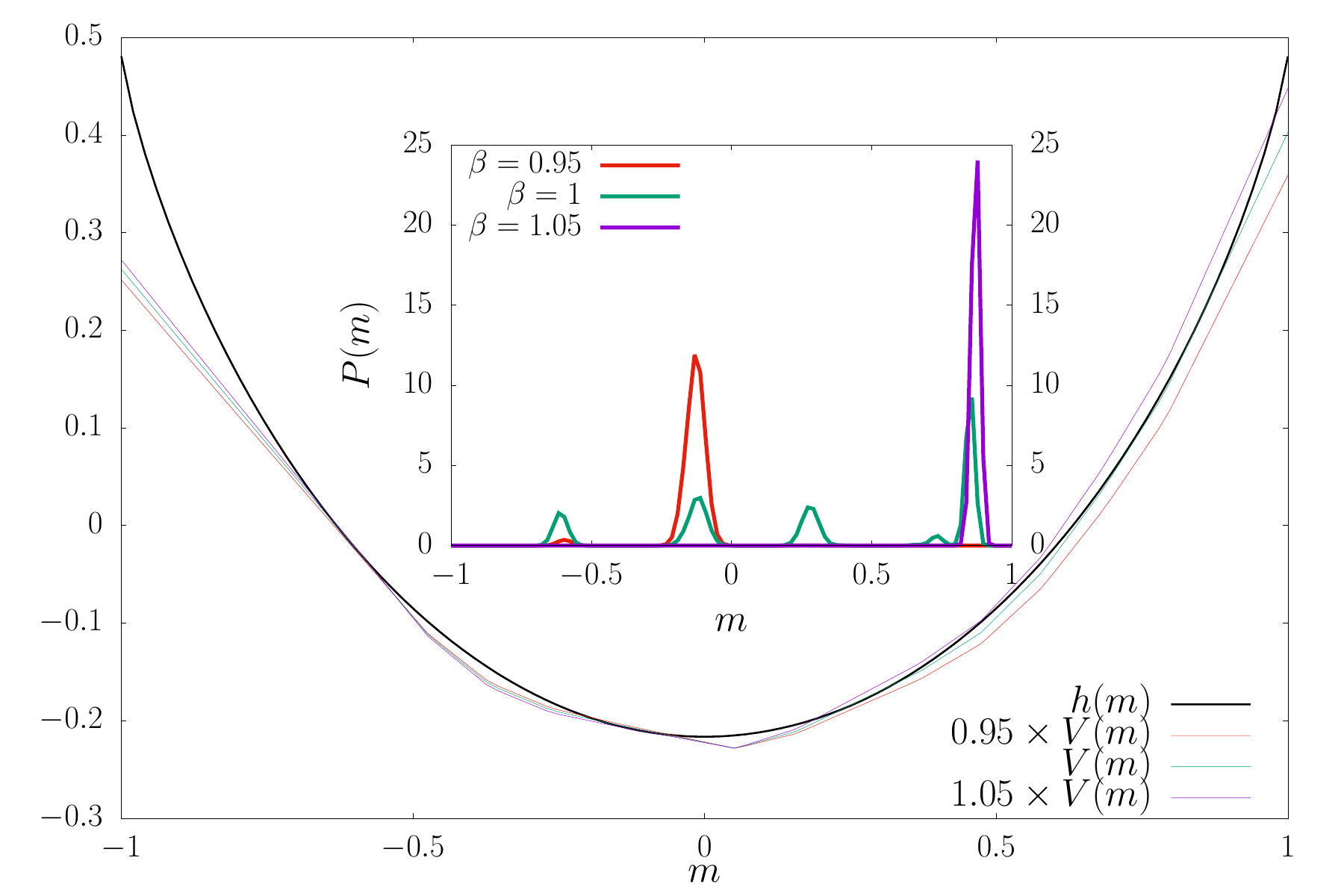}
\caption{First order phase transition mechanism of the ``Coulomb'' RBM illustrated on the $1$-d case.
  $\beta$ represents an annealing inverse temperature, which effect is to multiply the energy part $V(m)$ in the
  free energy $\F_\beta(m) = \beta V(m)-h(m)$. $\beta=1$ is a reference RBM where many states are present with comparable probabilistic weights;
  for $\beta>1$ [resp. $\beta<1$] the state with highest [resp. lowest] $V(m)$ is favored. \label{fig:rbm-beta}}
\end{figure}

In the $2$-d study case we have $\n_j = \bigl(\cos(p\pi/\sqrt{N_h}),\sin(p\pi/\sqrt{N_h})\bigr)$ and $z_j = -1+2r/\sqrt{N_h}$ with the entire decomposition of $j=p\sqrt{N_h}+r$.
The continuous dynamics of the parameters corresponding to an ordinary  gradient reads
\[
\dot \bq = \gamma\nabla_{\bq} \L[\bq],
\]
with the learning rate $\gamma$.
The scores associated with these parameters are centered random variables classically defined as
\[
\bq^*(\m) = \nabla_{\bq} \log p(\m\vert\bq),
\] 
which in our case read:
\[
q_j^*(\m) = \vert\n_j^T\m-z_j\vert - {\mathbb E}_{\m\sim p(\m\vert\bq)}\bigl[\vert\n_j^T\m-z_j\vert\bigr].
\]
In terms of these variables the gradient of the log likelihood simply reads
\[
\nabla_{\bq}\L[\bq] = {\mathbb E}_{\m\sim\hat p}\bigl[\bq^*(\m)\bigr],
\]
$\hat p$ being the empirical data distribution. In a continuous limit of the training process indexed by the training time $t$,
let $\hat \bq_t^*$ the time dependent expectation of $\bq_t(\m)$
w.r.t the empirical distribution $\hat p$. We have 
\[
\hat q_{j,t}^* =  \int d\m\bigl(\hat p(\m)-p(\m\vert\bq_t)\bigr) \vert \n_j^T\m-z_j\vert.
\]
The evolution of  $\hat \bq_t^*$ with time is given by
\[
\frac{d}{dt}\hat \bq_t^* = -\gamma \Cov_{\m\sim p(\m\vert\bq)}\bigl[\bq_t^*(\m),\bq_t^{*T}(\m)\bigr]
\hat \bq_t^*,
\]
where $\Cov_{\m\sim p(\m\vert\bq)}$ denotes the covariance under $p(\m\vert\bq)$ and corresponds to the Fisher metric of the $\bq$ parameter space. 
As we see as long as the covariance matrix is strictly positive definite the dynamics is contractant. When using the natural gradient~\cite{Amari}
defined here as
\[
\tilde\nabla_{\bq} = \Cov_{\m\sim p(\m\vert\bq_t)}\bigl[\bq_t^*(\m),\bq_t^{*T}(\m)\bigr]^{-1} \nabla_{\bq},
\] 
the dynamics simplifies to
\[
\frac{d}{dt}\hat \bq_t^* = -\gamma \hat \bq_t^*.
\]
The norm of $\hat \bq_t^*$ can be monitored during learning allowing for an adaptive strategy for the learning rate used in practice in these experiments to control
the convergence as well as the stopping criterion.
\begin{figure}[ht]
\includegraphics[width=\columnwidth]{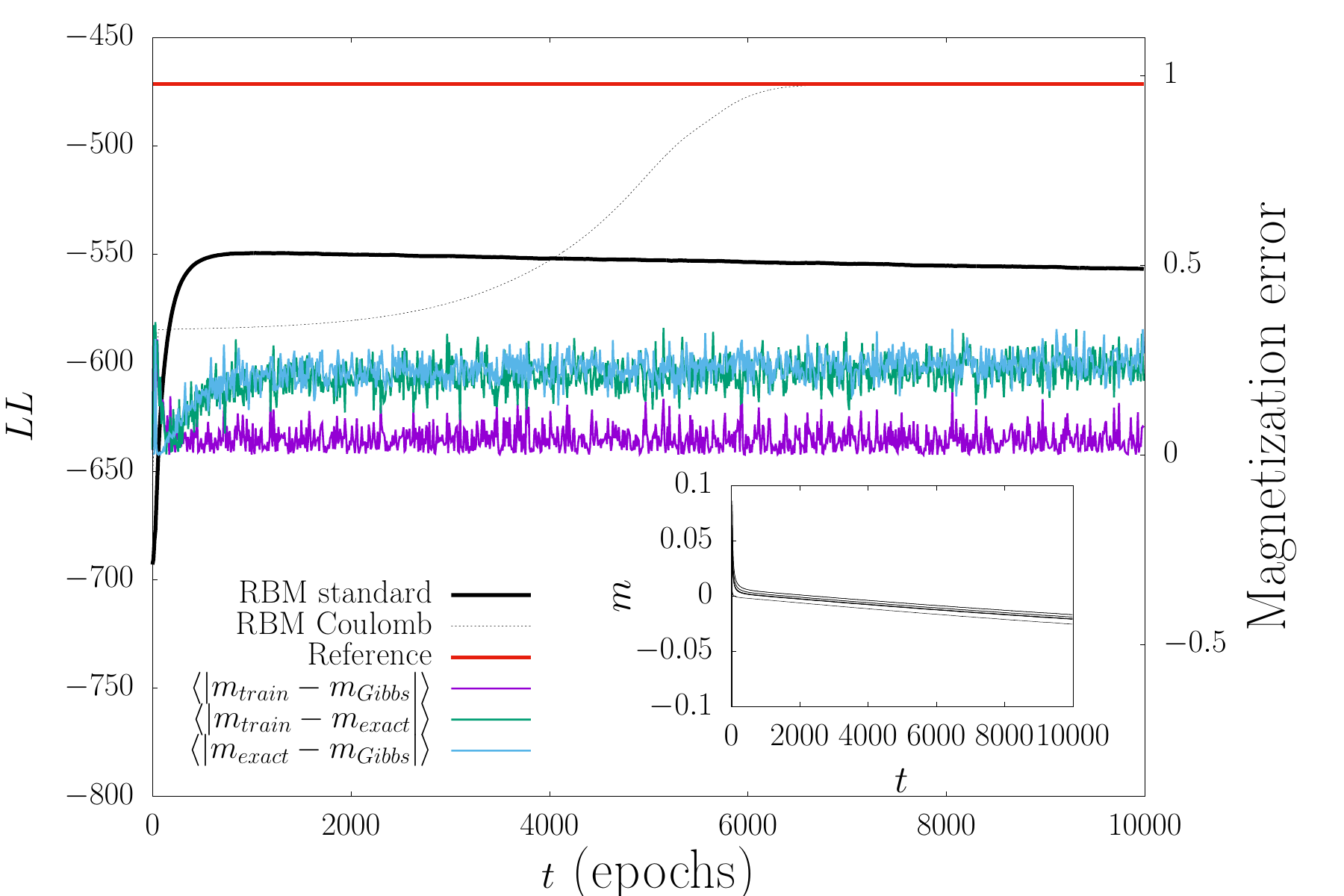}
\caption{Illustration of the failure of the standard RBM training due the occurrence of a first order transition signaled here by a divergence between the Gibbs sampling
  evaluation of the magnetization along the intrinsic axis with the  exact one (main plot), and due to the trapping of the $z_j$'s close to zero (inset).\label{fig:rbm-fail}}
\end{figure}
Thanks to the choice made for $u_\alpha$ in these experiments the number of configurations $\m$ is equal to $N+1$ and $(N+1)^2/4$ respectively in the $1$ and $2$-d case,
so the LL of the resulting ``Coulomb'' RBM which are  obtained can be evaluated exactly
in both cases as well as the corresponding standard RBM. The latter is obtained through the following asymptotic
mapping assuming  that $\sqrt{\sum_i W_{ij}^2}$ is large for all $j$:
\[
\log\cosh\bigl(\sum_{i=1}^{N_v}W_{ij}s_i - \theta_j\bigr) \approx \sqrt{N_v\sum_i W_{ij}^2}\Bigl\vert \frac{\bs^T}{\sqrt{N_v}}\n_j-z_j\Bigr\vert
\]
with
\[
  n_{ij} = \frac{W_{ij}}{\sqrt{\sum_i W_{ij}^2}}\qquad\text{and}\qquad
  z_j = \frac{\theta_j}{\sqrt{N_v\sum_i W_{ij}^2}}
\]
leading to
\[
q_j = \sqrt{\frac{1}{N_v}\sum_i W_{ij}^2}.
\]
This results into the following correspondence
\[
 W_{ij} = \sqrt{N_v}q_j n_{ij},\qquad 
 \theta_j = N_v q_j z_j.
 \vspace{0.2cm}
\]
In both experiments the number of visible variables is $N_v=10^3$, and the LL is an average value estimated on an independent test set of $10^3$ samples.
The number of points $N_{\rm{pts}}^d$ used to estimate the integrals over $\m$ needed to compute the natural 
gradient give a contribution $N_f N_{\rm{pts}}^d$ to the complexity in a naive setting.
The indicated values in the Table~\ref{tab:LLscores} correspond to the point where the results become insensitive to $N_{\rm{pts}}^d$. 
The values $LL_{\rm{Coulomb}}$ and $LL_{\rm{RBM}}$ measured 
respectively for the ``Coulomb" machine which is optimized with the natural gradient and its corresponding RBM using the previous mapping are reported on Table~\ref{tab:LLscores} are
compared with the reference value $LL_{\rm{Ref}}$ of the hidden mixture model used to generate the data.
Note that the mapping gives a poor model when many weak features are used as in the $d=2$ case with $N_h=900$.
Note also that using the ordinary gradient instead of the natural one seems to keep the last bits of LL out of reach in a reasonable time.

Finally Fig.~\ref{fig:rbm-fail}
illustrates the reasons for the failure of the standard RBM training. First  the Gibbs sampling procedure is plagued by the presence of $1$st order phase transitions
which is well understood when considering the ``Coulomb" RBM. Indeed, in that case changing the
temperature corresponds to multiplying all the feature weights $q_j$ by a common factor $\beta$ representing for instance an annealing inverse temperature. The learning procedure is supposed to tune
precisely the difference $\F_\beta[\m\vert\bq] = \beta V(\m\vert\bq)-\Sc[\m]$ at $\beta=1$, in order to obtain many coexisting states  corresponding to different values of condensed magnetization $\m$.
Then as in the example of Fig.~\ref{fig:rbm-beta}, changing slightly $\beta$ has the effect of concentrating the probability distribution on the state with highest or
lowest value of $V(\m\vert\bq)$ depending on whether $\beta$ is smaller or greater than one. The second source of failure is, as expected from  the electrostatic picture,
that the hidden bias given in rescaled form by $z_j=\theta_j/\nu_j$
along with~(\ref{eq:nuj}) get trapped, around zero in the example of Fig.~\ref{fig:rbm-fail} which prevent the machine to form more than two ferromagnetic states in $1$d.

\end{appendix}

\end{document}